# Effect of matrix sparsity and quantum noise on quantum random walk linear solvers

*Benjamin Wu[1], Hrushikesh Patil[2], Predrag Krstic[3]*

[1]Department of Applied Mathematics and Statistics, Stony Brook University, Stony Brook NY 11794

[2]Department of Electrical and Computer Engineering, North Carolina State University, Raleigh NC 27606

[3]Institute for Advanced Computational Science, Stony Brook University, Stony Brook NY 11794-5250

**Abstract**

We study the effects of quantum noise in hybrid quantum-classical solver for sparse systems of linear equations using quantum random walks, applied to stoquastic Hamiltonian matrices. In an ideal noiseless quantum computer, sparse matrices achieve solution vectors with lower relative error than dense matrices. However, we find quantum noise reverses this effect, with overall error increasing as sparsity increases. We identify invalid quantum random walks as the cause of this increased error and propose a revised linear solver algorithm which improves accuracy by mitigating these invalid walks.

## I. INTRODUCTION

Solving systems of linear equations Ax=b (SLE) is useful for many applications of computational science. This includes linear programming, the PageRank search indexing algorithm, and problems in machine learning such as linear regression [1-3]. Quantum computing algorithms have been developed for solving SLE, notably the HHL algorithm [4]. However, traditional quantum algorithms require deep quantum circuits and are limited by current noisy intermediate-scale quantum (NISQ) devices, which are subject to depolarization, decoherence, and measurement error [5]. Thus, hybrid-quantum classical algorithms such as the variational quantum linear solver (VQLS) algorithm [6-9] and the quantum random walk (QRW) linear solver [10] were proposed to shorten circuit depth and achieve noise resilience [11].

A matrix inversion method proposed by Ulam and Von Neumann uses random walks to solve SLE [12-14]. Chen et al. [10] examine this random walk linear solver for a case of stoquastic matrices

$$\boldsymbol{A} = 1 - \gamma \boldsymbol{P} \qquad (1)$$

where $\boldsymbol{P}$ defines a stochastic matrix and $\gamma$ is a real number such that $0 < \gamma < 1$

The solution of these SLE can be expressed in terms of a truncated Neumann series expansion, with truncation variable $c$. The component $I_0$ of the solution vector $x = \boldsymbol{A}^{-1}b$ is written in the form

$$x_{I_0}^{(c)} = \sum_{s=0}^{c} \gamma^s \sum_{I_1,\ldots,I_s=0}^{N-1} P_{I_0,I_1} \ldots P_{I_{s-1},I_s} b_{I_s} \qquad (2)$$

The classical Ulam-von Neumann algorithm computes this truncated series by using Monte Carlo random walks [12-14]. Every element in the solution vector is assigned to a numbered node in the random walk, and the transition probability from node $i$ to node $j$ is $P_{i,j}$, i.e., the element in the $i$-th row and $j$-th column of **P**. To calculate component $I_0$ of the solution vector, a random walk starts on the random walk node corresponding to $\boldsymbol{b}_{I_0}$, and each step is sampled randomly from the transition probability matrix $P$. The value of $c$ required to achieve a desired sampling error $\epsilon$ can be estimated with

$$c \sim \log(1/\epsilon)/\log(1/\gamma). \tag{3}$$

We investigate the QRW linear solver under the effect of quantum noise in the case of sparse matrices. SLE algorithms such as the classical conjugate gradient method and the quantum HHL algorithm have a known complexity relation to the sparsity of matrices. However, there is an absence in the literature on how the QRW linear solver algorithm behaves for sparse matrices. We find that accuracy of the linear solver improves with higher sparsity in ideal noiseless conditions, but the reverse is true in noisy conditions. This is because quantum noise leads to invalid quantum random walks for sparse matrices, resulting in device noise-based error in addition to baseline sampling error as defined in (3). By mitigating these invalid random walks with a simple "detect-and retry" approach, we recover the expected relation between sparsity and accuracy of the QRW linear solver algorithm in presence of quantum noise.

In section II, we address methods for executing the QRW linear solver on a quantum circuit, formulating sparse matrices, and executing the solver algorithm on simulated quantum computer backends. In section III, we present results for the algorithm on noiseless and noisy simulated systems, investigate the effect of quantum noise on the QRW linear solver, and propose an adjusted algorithm to mitigate the effect of noise. In section IV, we summarize the effectiveness of our mitigation approach.

## II. METHODS

A special case of random walk matrices with dimensions $N \times N = 2^n \times 2^n$ represented by a Hamming Cube [10] enables the linear solver in (Eq. 2) to be executed on a quantum computer. Each node is represented as a binary string of length $n$, and transition probabilities are only defined across nodes that have a Hamming distance of 1, as shown in figure 1. The probability of transitioning across other node pairs is the product of the intermediary transition probabilities. Equivalently, to decide a next step, a quantum "coin" is tossed for each of the $n$ binary string elements to decide if it will flip or stay the same [15]. Each of the $n$ coins is represented by a single-qubit unitary operator, defined by angles $\theta, \phi, \lambda$:

$$U(\theta, \phi, \lambda) = \begin{bmatrix} \cos\left(\frac{\theta}{2}\right) & -e^{i\lambda}\sin\left(\frac{\theta}{2}\right) \\ e^{i\phi}\sin\left(\frac{\theta}{2}\right) & e^{i(\lambda+\phi)}\cos\left(\frac{\theta}{2}\right) \end{bmatrix}. \tag{4}$$

The transition probability between start node $J'$ and end node $J$ in stochastic matrix $\mathbf{P}$ is defined as

$$P^{quantum}_{J',J} = \prod_{l=0}^{n-1} |U(\theta_l, \phi_l, \lambda_l)_{i_l, i_{l-1}}|^2 \tag{5}$$

with binary string $|I\rangle = |i_{n-1}, \dots, i_1, i_0\rangle = |J'\rangle \oplus |J\rangle$ [10]. We note that the absolute value in (Eq. 5) removes the effect of angles $\phi$ and $\lambda$ on the probabilities. Thus, the Hamming Cube matrices are defined in terms of $\theta$ from this point onwards.

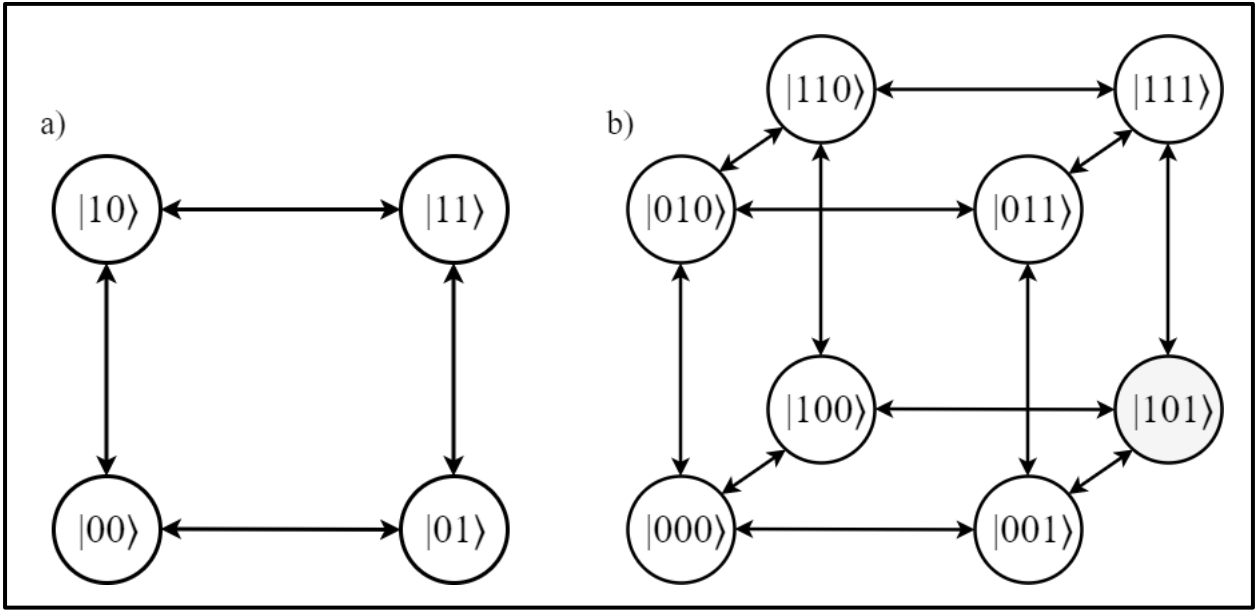

**FIG. 1.** Transition graph diagrams for the Hamming Cube random walk for a) n=2 and b) n=3. Nodes that have a Hamming distance of 1 are connected by undirected edges

The equivalent quantum circuit applies the unitary gate corresponding to each θ value, followed by a CNOT gate to create a correlated structure, as shown in figure 2. A single quantum random walk step is completed when the circuit is measured, and the resulting values are used to initialize the next step. Fig. 2 illustrates the $n = 2$ and n = 3 cases.

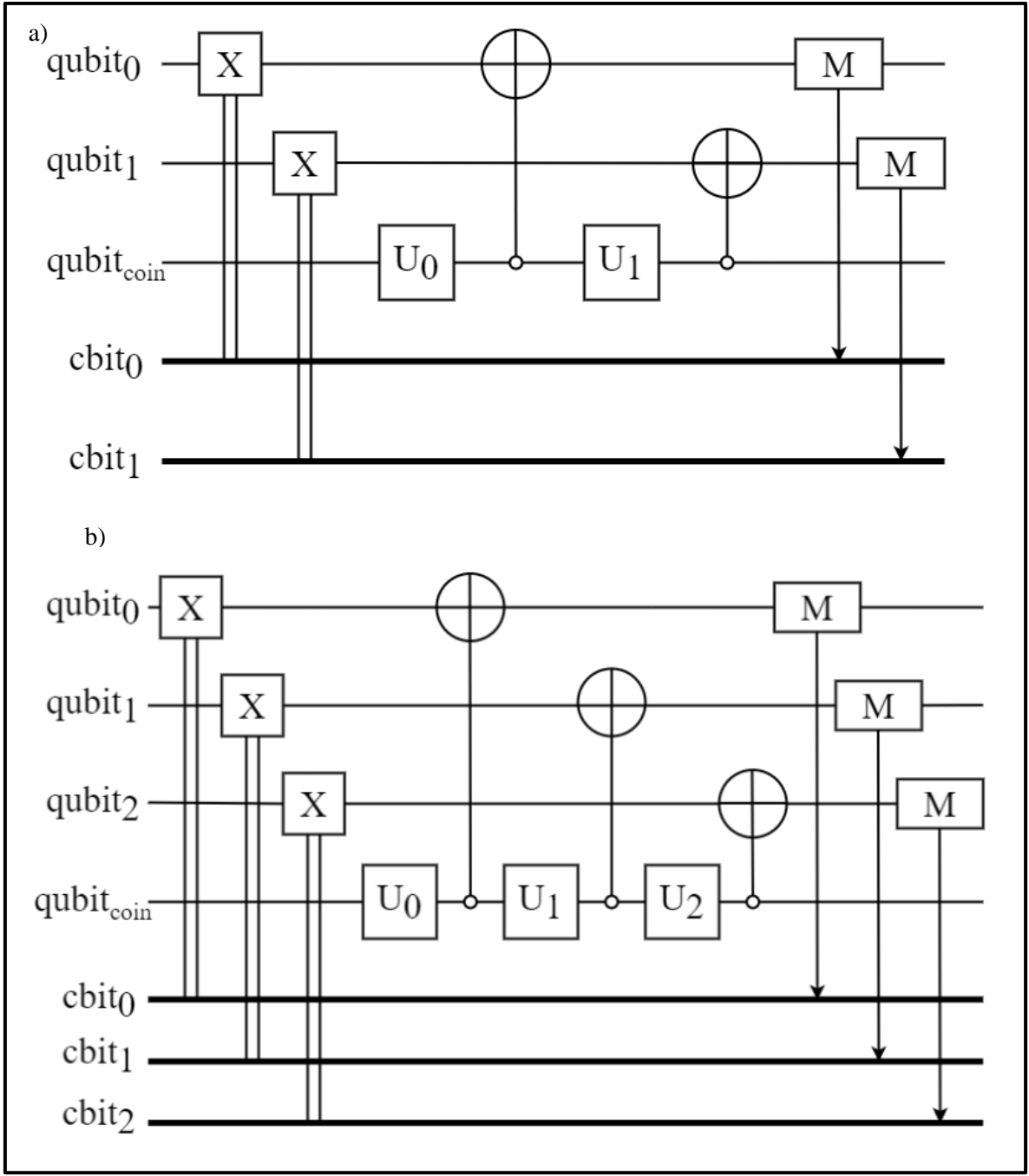

**FIG. 1**. Quantum random walk step circuit for a) N=4 (n=2) and b) N=8 (n=3), using a paired unitary-CNOT operation for each coin. Following execution of each quantum circuit, the results of measurement initialize the state for the next step via classical conditional gates [10].

We conduct tests for matrices of dimensions N=4 (n=2), N=8 (n=3), and N=16 (n=4). For each matrix size, values of $\theta$ were generated uniformly randomly in the range $[-\pi,\pi]$ to define a dense matrix. Then matrices of increasing sparsity were obtained by setting each $\theta$ to 0. Our definition of sparsity for a given matrix is the fraction of the number of zero elements over the total number of elements [9]. For N=4, we used sparsity levels of 0, 0.5, and 0.75. For N=8, we used sparsity levels of 0, 0.5, 0.75, 0.875. For each case of matrix size and sparsity level, 50 matrix samples of varying condition numbers were generated in this fashion. For matrices of the same size, an Nx1 vector **b** was generated randomly from a uniform distribution in the range [-1, 1]. Further details are provided in the supplementary information (SI).

The code for the QRW algorithm was written in Python using the IBM Qiskit package and executed on simulated IBM-Quantum backends [16]. For the ideal noiseless case we used the noiseless QASM simulator. For the noisy case we used the QASM simulator with the Qiskit noise models, which recreate the noise parameters of the legacy IBM-Q Boeblingen 5-qubit and Casablanca 7-qubit backends, as reported in the SI. Multiple shots of the QRW algorithm are required to obtain an accurate solution vector, thus for each test the number of shots was varied geometrically from 24 to 1008. The shot counts were chosen to be multiples of 24, keeping in mind the number of cores on the simulation cluster. The solution vector resulting from each execution was then compared to the standard numerical solution computed using NumPy to get the relative error. The relative error was defined by the L-2 norm of the difference between the two solutions, divided by the L-2 norm of the numerically calculated solution $A^{-1}b$. Graphs including standard error are shown in the results section.

**III. RESULTS**

Our results in figure 3 show that on the noiseless simulated quantum computer, the QRW algorithm attains lower relative error as the sparsity of the matrices increases, as expected. Opposite to the noiseless case, the QRW algorithm on the noisy quantum simulator performs *worse* as the sparsity of the matrices increases, as shown in figure 4. For the same number of shots in the algorithm, sparse matrices achieved consistently higher errors than denser counterparts, and this was true for N=4, N=8, and N=16 systems. Equivalently, less shots were required by sparse matrices to achieve the same level of error than dense matrices under noisy conditions. This result is the reverse of the expected behavior regarding matrices of decreasing condition numbers and contradicts the results from the noiseless case.

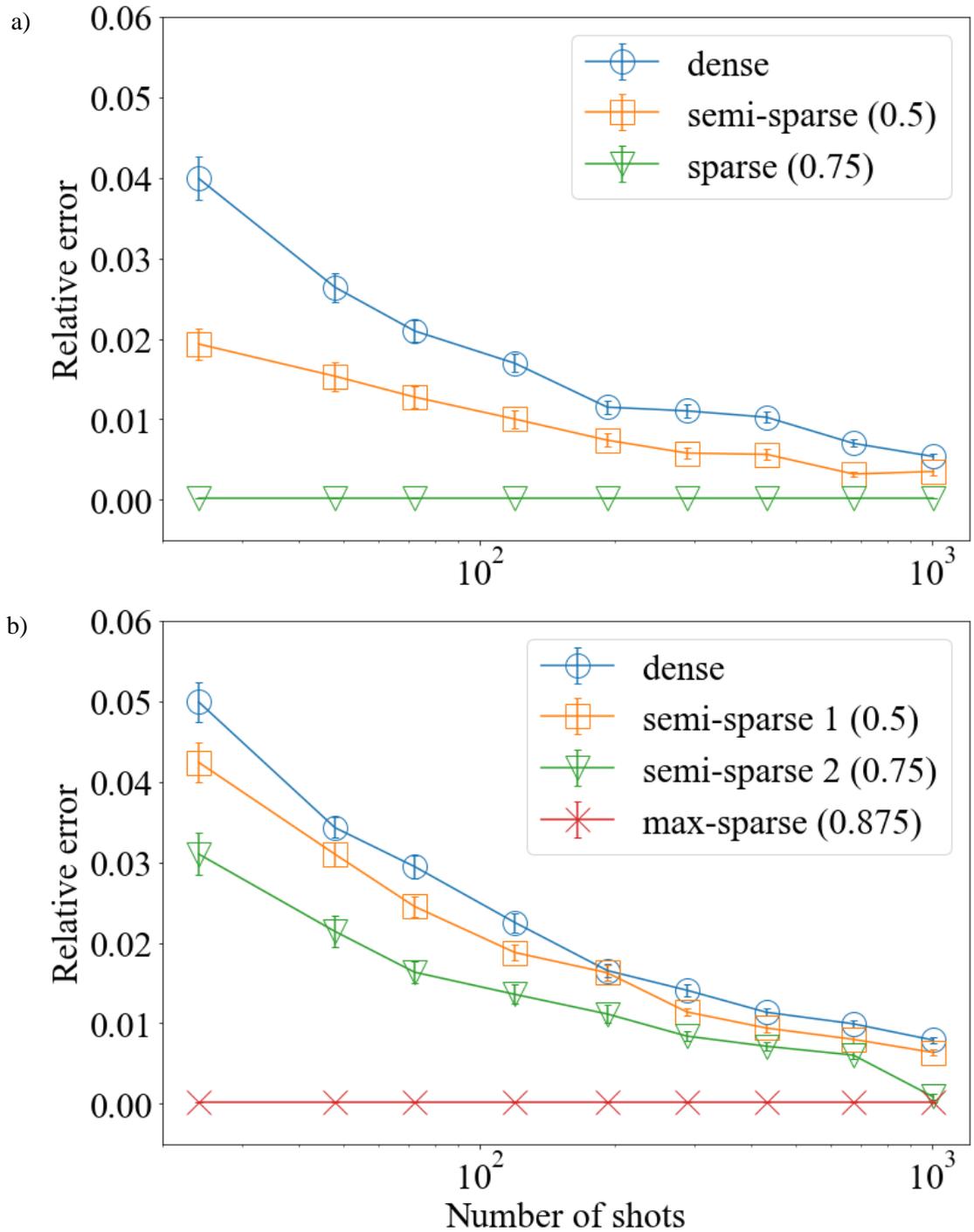

**FIG. 2**. Ideal noiseless simulations using the QASM backend for: a) 4x4 matrix with 3 sparsity levels; b) 8x8 matrix with 4 sparsity levels. Shots are scaled logarithmically on the x-axis.

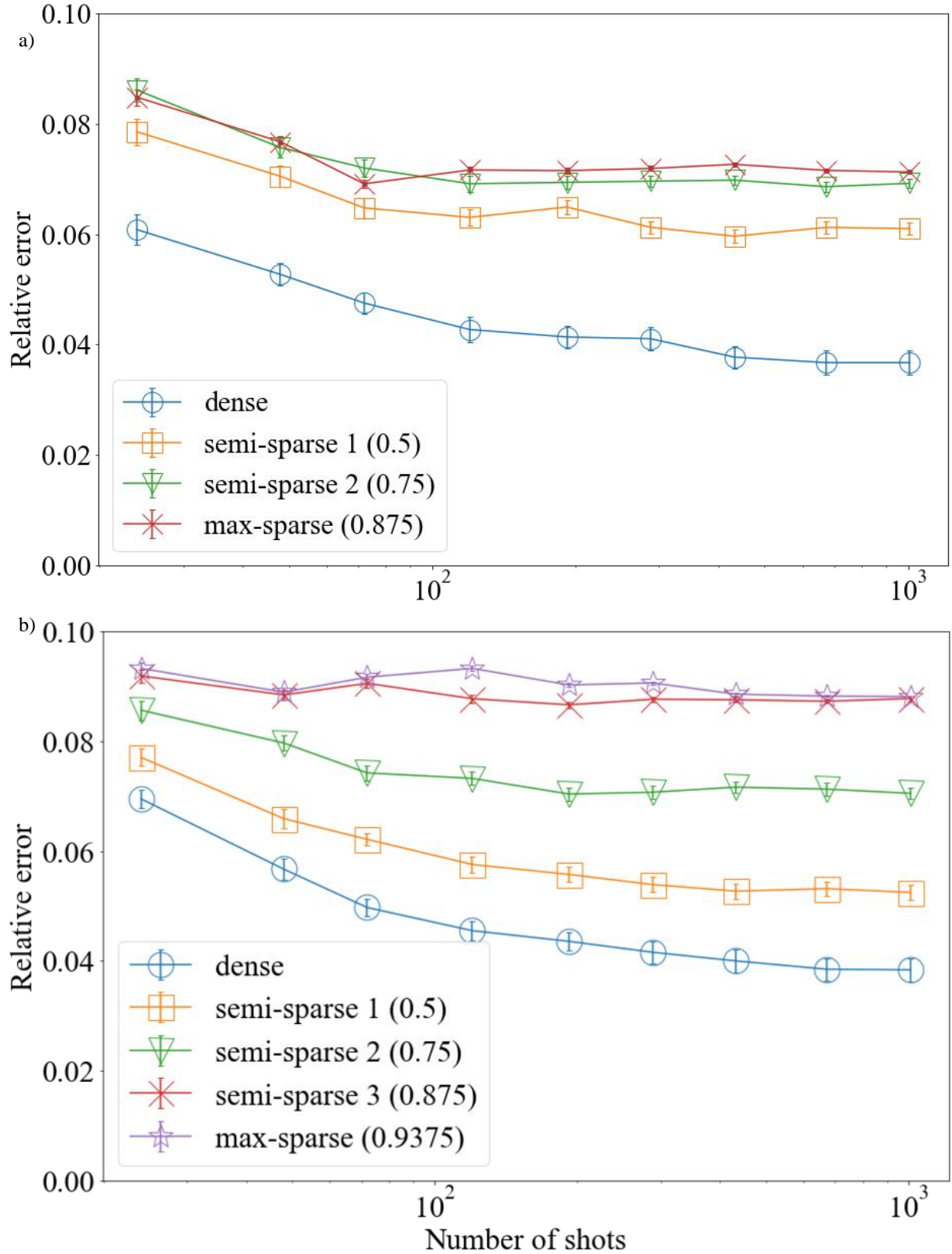

**FIG. 4.** Simulations of the Fake Boeblingen noise model for a) N=8 matrix with 4 sparsity levels, and b) N=16 matrix with 5 sparsity levels. Shots are displayed on the x-axis with a logarithmic scale.

We show how quantum noise and sparsity lead to increased errors. For each zero value of $\theta$ in a sparse matrix, a transition probability of zero exists somewhere in the random walk. If a step with zero probability in the transition matrix occurs due to quantum noise, we call it an *invalid step*. In an ideal noiseless case, a transition probability of zero behaves as expected and it is impossible for an invalid step to occur. However, in the presence of noise, quantum error may lead to a step even when the transition probability is zero, creating an invalid step. If such a step occurs, the remaining random walk diverges from the expected random walk corresponding to (Eq. 2). As a result, the error caused by an invalid step is greater than the expected truncation error caused by a valid step. In figure 5 we show that matrices with higher sparsity result in more invalid steps, and in figure 6a this leads to higher relative error.

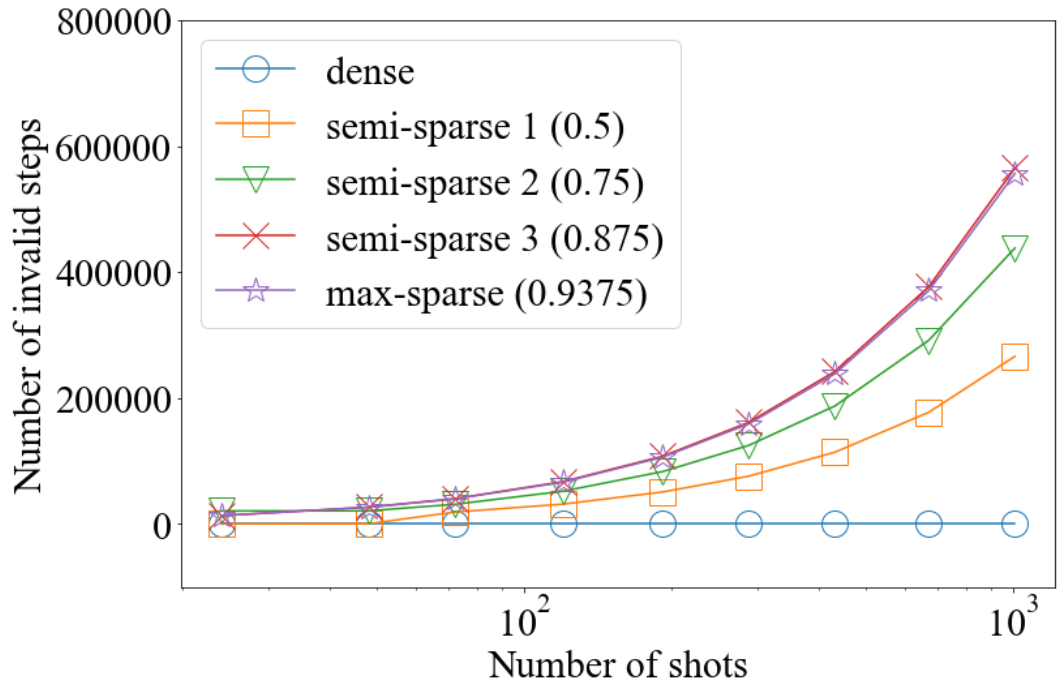

**FIG. 5.** Plot of invalid step count versus number of quantum random walk shots for N=16 on the Fake Casablanca backend. The number of invalid steps increases as sparsity of system increases.

To mitigate the effect of invalid random walks, we devise an adjusted algorithm that detects and retries an invalid step and retries until it is valid. Given current node $i$ and adjacent node $j$ from the quantum circuit measurement, we identify the step $i$-to-$j$ as invalid if the transition matrix element $P_{i,j} = 0$. This detection subroutine uses a simple array lookup and requires only O(1) time complexity per step. We then retry the quantum circuit and measure a new adjacent node $j'$ until $P_{i,j'}$ is non-zero.

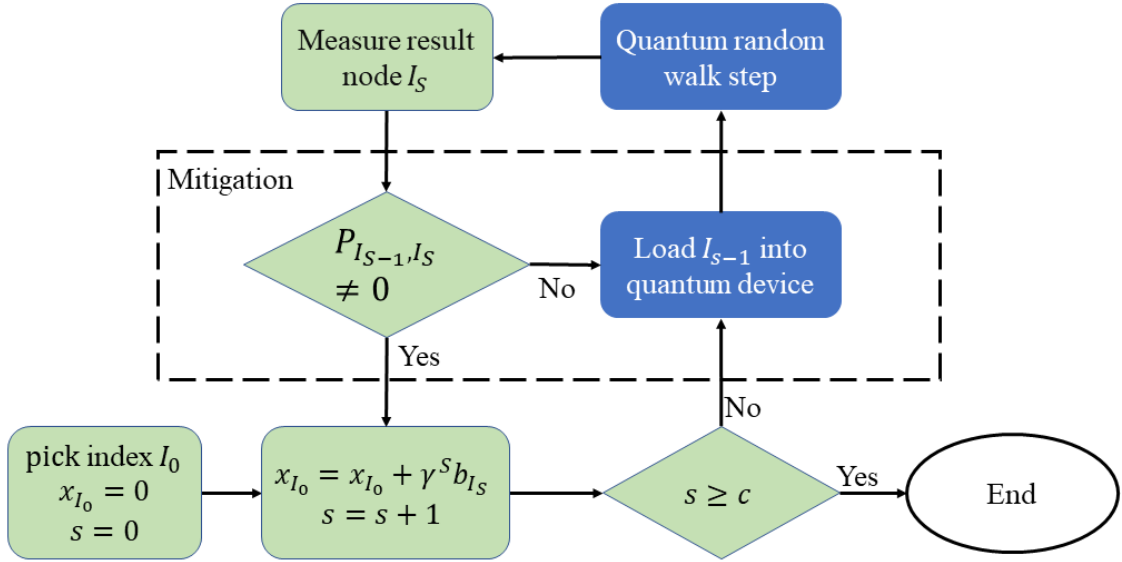

**Algorithm 1**: Classical procedures (light green) include storing solution vector component, measurement, and validation; quantum procedures (dark blue) include loading state and performing random walk. The proposed mitigation procedure is denoted by a dotted line.

We compare the QRW solver performance with and without this mitigation adjustment on the simulated Casablanca backend. In figure 6a, we show the default algorithm achieves worsening accuracy for increasingly sparse matrices, which corroborates the results on the Boeblingen backend in figure 4 and contradicts the expected ideal behavior. From our previous analysis, we deduce this is because the default algorithm does not distinguish between invalid and valid random walks.

|  | *Matrix Sparsity Level* | | | | |
| --- | --- | --- | --- | --- | --- |
|  | *0 (dense)* | *0.50* | *0.75* | *0.875* | *0.9375* |
| *Un-mitigated* | *3.71* | *4.34* | *6.35* | *8.41* | *8.44* |
| *Mitigated* | *3.66* | *2.59* | *1.91* | *0.56* | *0.02* |

**Table 1:** Relative error (%) comparison, for N=16, with 1008 shots

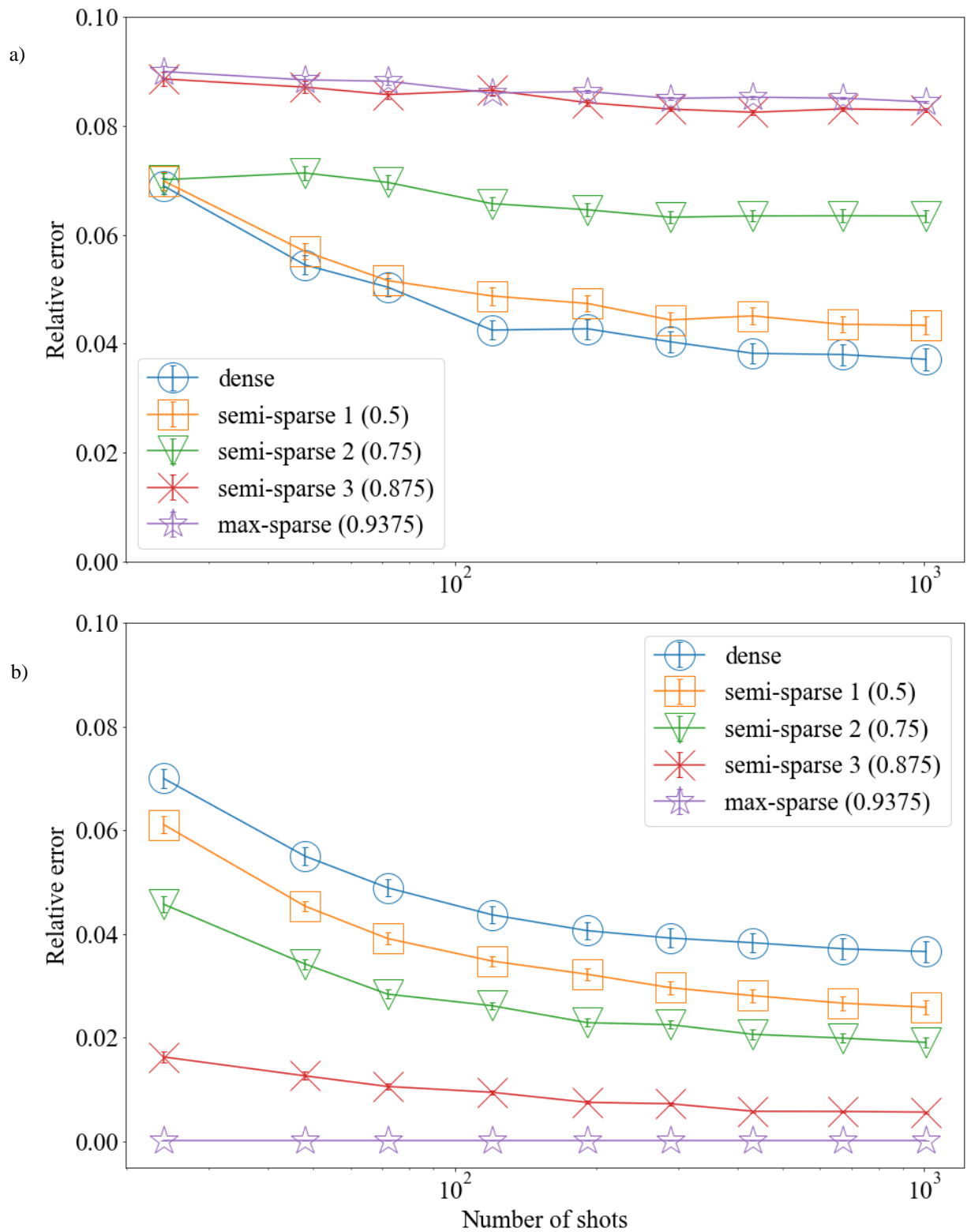

**FIG. 6**. Noisy simulations using the Fake Casablanca noise model for algorithm: a) without mitigated invalid steps, b) with mitigated invalid steps.

With the invalid walk mitigation adjustment, the expected that behavior of the algorithm is successfully recovered. In figure 6b, we show that the mitigated algorithm achieves lower accuracy for increasingly sparse matrices, which mirrors the noiseless case in figure 3. Accordingly, the mitigation improves accuracy for all levels of sparsity, as shown in table 1.

## IV. CONCLUSION

In this work we investigate the effect of the quantum noise on the quantum random walk linear solver with sparse matrices. In the ideal noiseless case, the QRW linear solver achieves lower sampling error for matrices of increasing sparsity. However, we show that quantum noise counter-intuitively reverses this relation between sparsity and sampling error. In the presence of noise, invalid random walks negatively affect the accuracy of the algorithm on sparse matrices corresponding to steps with zero probability. To mitigate the effect of invalid walks, we propose a simple detect-and-retry fix. This adjustment recovers the expected behavior of the algorithm in relation to sparsity, and improves accuracy compared to the default implementation.

Besides the QRW linear solver, there are other algorithms that make use of quantum random walks, such as graph traversal and triangle finding [17,18]. In the case that these algorithms contain zero transition probabilities and experience invalid random walks under noisy conditions, our approach for detecting and retrying random walks may be useful.

## V. ACKNOWLEDGEMENTS

The authors would like to thank Stony Brook Research Computing and Cyberinfrastructure for access to the SeaWulf computing system, which was made possible by a $1.4M National Science Foundation grant (#1531492). In addition, access to quantum computing systems for preliminary testing was provided by the IBM Quantum Hub at Oak Ridge National Laboratories

## VI. REFERENCES

[1] M. P. Deisenroth, A. A. Faisal, and C. S. Ong, "Linear Algebra," in *Mathematics for Machine Learning*, Cambridge: Cambridge University Press, (2020).

[2] H. M. Markowitz, "The elimination form of the inverse and its application to linear programming," *Management Science* **3**, 255, (1957).

[3] A. Langville and C. Meyer, "Deeper inside PageRank," *Internet Mathematics*, **1**, 335, (2004).

[4] Harrow, Aram W. et al. Quantum Algorithm for Linear Systems of Equations. Physical Review Letters **103**, (2009).

[5] K. Georgopoulos, C. Emary, and P. Zuliani, "Modeling and simulating the noisy behavior of near-term quantum computers," *Physical Review A*, **104**, (2021).

[6] C. Bravo-Prieto, R. LaRose, M. Cerezo, Y. Subasi, L. Cincio, and P. J. Coles, Variational quantum linear solver (2020), arXiv:1909.05820.

[7] H.-Y. Huang, K. Bharti, P. Rebentrost, 'Near-term quantum algorithms for linear systems of equations'. arXiv (2019).

[8] X. Xu, J. Sun, S. Endo, Y. Li, S. C. Benjamin, X. Yuan, 'Variational algorithms for linear algebra', *Science Bulletin*, **66** (2021).

[9] H. Patil, Y. Wang, and P. S. Krstić, "Variational quantum linear solver with a dynamic ansatz," *Physical Review A*, **105** (2022).

[10] Chen, CC., Shiau, SY., Wu, MF. et al. Hybrid classical-quantum linear solver using Noisy Intermediate-Scale Quantum machines. Sci Rep **9**, 16251 (2019).

[11] S. Endo, Z. Cai, S. C. Benjamin, en X. Yuan, "Hybrid Quantum-Classical Algorithms and Quantum Error Mitigation", *Journal of the Physical Society of Japan*, **90**, (2021).

[12] G. E. Forsythe and R. A. Leibler, "Matrix inversion by a Monte Carlo method," *Mathematics of Computation*, **4**, 127, (1950).

[13] Ji, Hao & Mascagni, Michael & Li, Yaohang, Convergence Analysis of Markov Chain Monte Carlo Linear Solvers Using Ulam--von Neumann Algorithm. SIAM Journal on Numerical Analysis. **51**, (2013).

[14] F. Xia, J. Liu, H. Nie, Y. Fu, L. Wan, en X. Kong, "Random Walks: A Review of Algorithms and Applications", *CoRR*, (2020).

[15] J. Kempe, "Quantum random walks: An introductory overview", *Contemporary Physics*, **44**, 307, (2003).

[16] G. Aleksandrowicz et al., Qiskit: An Open-source Framework for Quantum Computing (2019).

[17] F. L. Gall, "Improved quantum algorithm for Triangle finding via combinatorial arguments," *2014 IEEE 55th Annual Symposium on Foundations of Computer Science*, (2014).

[18] N. Shenvi, J. Kempe, and K. B. Whaley, "Quantum random-walk search algorithm," *Physical Review A*, **67**, (2003).

# Effect of matrix sparsity and quantum noise on quantum random walk linear solvers

*Benjamin Wu[1], Hrushikesh Patil[2], Predrag Krstic[3]*

[1]Department of Applied Mathematics and Statistics, Stony Brook University, Stony Brook NY 11794

[2]Department of Electrical and Computer Engineering, North Carolina State University, Raleigh NC 27606

[3]Institute for Advanced Computational Science, Stony Brook University, Stony Brook NY 11794-5250

*SI. Generating Sparse Matrices*

For a given system of equations of size $N \times N = 2^n \times 2^n$, first uniformly generate $n$ angle triplets for a given system of equations, where each angle is in the range $[-\pi, \pi]$. This yields a dense matrix. Then, for each increment of sparsity, one additional theta angle is changed from a nonzero to a zero value. This leads to sparsity levels $z$ occupying

$$1 - \left(\frac{1}{2}\right)^1, 1 - \left(\frac{1}{2}\right)^2, \ldots, 1 - \left(\frac{1}{2}\right)^{n-1}. \tag{S1}$$

For example, if we have $n = 2$, and angles $(\theta_0, \phi_0, \lambda_0), (\theta_1, \phi_1, \lambda_1)$, there are three levels of sparsity. The probability matrix for a dense case is:

$$P_{dense} = \begin{bmatrix} \cos^2(\frac{\theta_0}{2})\cos^2(\frac{\theta_1}{2}) & \sin^2(\frac{\theta_0}{2})\sin^2(\frac{\theta_1}{2}) & \cos^2(\frac{\theta_0}{2})\sin^2(\frac{\theta_1}{2}) & \sin^2(\frac{\theta_0}{2})\cos^2(\frac{\theta_1}{2}) \\ \sin^2(\frac{\theta_0}{2})\sin^2(\frac{\theta_1}{2}) & \cos^2(\frac{\theta_0}{2})\cos^2(\frac{\theta_1}{2}) & \sin^2(\frac{\theta_0}{2})\cos^2(\frac{\theta_1}{2}) & \cos^2(\frac{\theta_0}{2})\sin^2(\frac{\theta_1}{2}) \\ \cos^2(\frac{\theta_0}{2})\sin^2(\frac{\theta_1}{2}) & \sin^2(\frac{\theta_0}{2})\cos^2(\frac{\theta_1}{2}) & \cos^2(\frac{\theta_0}{2})\cos^2(\frac{\theta_1}{2}) & \sin^2(\frac{\theta_0}{2})\sin^2(\frac{\theta_1}{2}) \\ \sin^2(\frac{\theta_0}{2})\cos^2(\frac{\theta_1}{2}) & \cos^2(\frac{\theta_0}{2})\sin^2(\frac{\theta_1}{2}) & \sin^2(\frac{\theta_0}{2})\sin^2(\frac{\theta_1}{2}) & \cos^2(\frac{\theta_0}{2})\cos^2(\frac{\theta_1}{2}) \end{bmatrix} \tag{S2}$$

We get the first level of sparsity by setting the value of $\theta_0 = 0$:

$$P_{semi\_sparse} = \begin{bmatrix} 1 \cdot \cos^2\left(\frac{\theta_1}{2}\right) & 0 \cdot \sin^2(\frac{\theta_1}{2}) & 1 \cdot \sin^2(\frac{\theta_1}{2}) & 0 \cdot \cos^2\left(\frac{\theta_1}{2}\right) \\ 0 \cdot \sin^2(\frac{\theta_1}{2}) & 1 \cdot \cos^2(\frac{\theta_1}{2}) & 0 \cdot \cos^2\left(\frac{\theta_1}{2}\right) & 1 \cdot \sin^2(\frac{\theta_1}{2}) \\ 1 \cdot \sin^2(\frac{\theta_1}{2}) & 0 \cdot \cos^2\left(\frac{\theta_1}{2}\right) & 1 \cdot \cos^2(\frac{\theta_1}{2}) & 0 \cdot \sin^2(\frac{\theta_1}{2}) \\ 0 \cdot \cos^2\left(\frac{\theta_1}{2}\right) & 1 \cdot \sin^2(\frac{\theta_1}{2}) & 0 \cdot \sin^2(\frac{\theta_1}{2}) & 1 \cdot \cos^2(\frac{\theta_1}{2}) \end{bmatrix} \tag{S3}$$

We get the maximum level of sparsity by setting $\theta_0 = 0$ and $\theta_1 = 0$:

$$P_{max\_sparse} = \begin{bmatrix} 1 \cdot 1 & 0 \cdot 0 & 1 \cdot 0 & 0 \cdot 1 \\ 0 \cdot 0 & 1 \cdot 1 & 0 \cdot 1 & 1 \cdot 0 \\ 1 \cdot 0 & 0 \cdot 1 & 1 \cdot 1 & 0 \cdot 0 \\ 0 \cdot 1 & 1 \cdot 0 & 0 \cdot 0 & 1 \cdot 1 \end{bmatrix} = I \tag{S4}$$

*SII. Simulating Quantum Noise*

The simulated quantum computers were executed using the IBM Quantum built-in noise models for the Boeblingen and Casablanca backends. The noise models automatically include values for thermal relaxation errors according to parameters $T_1$ and $T_2$, as well as values for CNOT error and readout error.

|  | Avg. T1 ($\mu s$) | Avg. T2 ($\mu s$) | Avg. CNOT Error | Avg. Readout Error |
|---|---|---|---|---|
| Fake Boeblingen | 72.775 | 153.457 | 0.03211 | 0.05258 |
| Fake Casablanca | 89.968 | 85.496 | 0.01274 | 0.01898 |

*Table S1: Average noise model parameter values for each simulated backend*